\documentclass[twoside,fleqn]{article}
\usepackage{amssymb}
\usepackage{npb}
\usepackage{graphics}
\newcommand{\be}{\begin{equation}}
\newcommand{\ee}{\end{equation}}

\def\beq{\begin{eqnarray}}    
\def\eeq{\end{eqnarray}}      

\def\al{\alpha}
\def\be{\beta}

\def\La{\Lambda}

\def\si{\sigma}

\def\La{\Lambda}

\title{Cosmological constant, renormalization group and
Planck scale physics
}

\author{Ilya L. Shapiro \address{Departamento de Fisica, ICE,
Universidade Federal de Juiz de
Fora, MG, Brazil},
 {Joan Sol\`a} \address{Departament d'Estructura i Constituents de la
Mat\`eria, Universitat de Barcelona, and CER for Astrophysics,
Particle Physics and Cosmology, Diagonal 647, E-08028 Barcelona,
 Catalonia, Spain}
}

\begin{document}

\begin{abstract}
Starting from generic quantum effects at the Planck scale $M_P$,
we find that the renormalization group running of the cosmological
constant (CC) at low energies is possible if there is a smooth
decoupling of all massive particles from $M_P$ to the mass of the
lightest neutrino, $m_{\nu}$. We discuss the theoretical
implications of this running for the ``old'' and ``new''
cosmological constant problems. Interestingly enough, the CC
running implies a strong relationship between quantum field theory
and cosmology, which should be observable in  the near future in
experiments such as SNAP through the measurement of a cubic
redshift dependence of the CC.
\begin{flushright}
{{UB-ECM-PF-03/15, DF-UFJF/05-2003 }}
\end{flushright}
\end{abstract}

\maketitle

\section{Introduction}
The concept of vacuum is one of the most intriguing ones in modern
science, for it shows how the human views on the nature evolve
with time. In classical physics, the vacuum is just empty space or
the place prepared for the dynamics of particles. However, in
quantum mechanics and QFT the vacuum is full of interesting
phenomena such as creation and annihilation of virtual particles.
This notion of vacuum led to the exceptional success of QED. Here,
and also in the Standard Model (SM) of strong and electroweak
interactions, the effects of virtual particles manifest themselves
through quantum corrections to ordinary observables (e.g.
cross-sections). However, switching on the gravitational field,
one gets a chance to observe the energy and pressure of vacuum
itself, for it appears as a component of the Einstein equations:
that one called the cosmological constant (CC). According to the
recent supernova data\,\cite{SN}, the energy of vacuum is of the
order of the present critical density and hence this ``dark
energy'' dominates over the density of ordinary matter, radiation
and even over the density of the dark matter which is requested
by the astrophysical observations. Taking into account that all
the matter content of the Universe has been, probably, created
out of the vacuum during the reheating period after inflation, the
non-zero energy of vacuum can not be seen as a total surprise. In
actual fact the most shocking issue is why there is nowadays a
residual vacuum energy (the observed CC)  so close to the matter
density. Furthermore, there is another, even greater, mystery in
this story. The naive estimate of the tree-level contribution of
the energy of vacuum in the SM (induced CC) is some $55$ orders of
magnitude greater than the cosmological constant which has been
detected via the supernovae observations. In order to cure this
problem one has to introduce another CC which is a characteristic
of the vacuum itself (vacuum CC). Then one is forced to fine-tune
this independent parameter with a tremendous precision. To explain
this fine-tuning in a natural way is the ``old'' CC problem
\cite{weinRMP}, while to explain the approximate coincidence of
the CC and the matter density is the ``new'' CC problem
\cite{weinDM}. Here we wish to address the ``new'' CC problem in
the light of the Renormalization Group (RG) in QFT in a curved
background. \vskip 6mm

\section{Renormalization group and decoupling}

In the standard picture the CC does not change when the Universe
expands in the FLRW phase, while the matter-radiation density is
rapidly decreasing. The only changes of the CC are associated to
phase transitions in the early Universe and should be constant in
the later epochs\,\footnote{Notice that every phase transition
should change the induced CC and leave the vacuum CC intact. That
is why the fine-tuning is so weird!}. But there is another
possible source of time dependence of the CC. Both induced and
vacuum CC are subject to the RG running\,\cite{nova}. However,
this running must be suppressed at low energies because it is
produced by the Feynman diagrams with loops of massive matter
fields with external gravitational tails. If we use Einstein
equations to estimate the typical energy of the gravitational
quanta, this energy must be associated to the Hubble parameter
$\,H$\,\cite{nova}. In the present universe this parameter has an
approximate value $\,H_0\sim 1.5\times 10^{-42}\,GeV$. If we
compare this number with the assumed mass of the presumably
lightest neutrino, $\,m_\nu \sim 10^{-12}\,GeV$, at first sight
it is clear that the corresponding loop must be completely
decoupled and the running of the CC at low energy has no sense.
But this is not the whole story. Consider the contribution of the
particle with mass $\,m\,$ to the $\,\be_\La$-function. The
decoupling of massive particles at low energies is not
abrupt\,\cite{Manohar}, and the usual form of the scale
dependence of this $\,\be$-function at $\,p^2\ll m^2\,$ (here
$\,p^2\,$ is the square of the Euclidean momenta) is expected to
be\,\cite{babic} \beq \be_\La^{(IR)} \,\sim\,
\frac{p^2}{m^2}\,\times\,{\rm const}\,. \label{m} \eeq Despite
that a recent attempt\,\cite{apco} to verify the
Appelquist-Carazzone theorem\,\cite{AC} in curved space-time
failed in the part concerning the CC, we have a very strong
argument in favor of the formula (\ref{m}). In order to see the
decoupling one has to apply the physical mass-dependent
renormalization scheme, which requires an explicit control of the
energy of the particles\,\cite{Manohar}. For this reason, the
calculations can be performed straightforwardly only for the
metric perturbations on the flat background, and this restricts
too much the form of the covariant non-local terms in the vacuum
effective action which may be responsible for the running of the
CC. The root of the problem is that we do not possess, at the
moment, a completely covariant calculational technique compatible
with the mass-dependent renormalization scheme. However, let us
suppose that we have the instruments for this calculation. Then,
in the cosmological setting, we are going to meet a general
$\,H$-dependent expression for the contribution of the particle
with mass $\,m$ and spin $J$:
\begin{equation}
\be_\La(m,H) \,=\,
F\Big(\frac{H^2}{m^2}\Big)\,\times\,\be_\La^{(UV)}\,, \label{beta
function}
\end{equation}
where $\be_\La^{(UV)}=N_J{m^4}/{16\pi^2}$ is the non-suppressed UV
contribution to the CC $\beta$-function\,\cite{nova}. Here
$N_J=(-1)^{2J}(J+1/2)\,n_J\,n_c$ is a multiplicity factor, with
$n_{\{0,1,1/2\}}=(1,1,2)$ and  $n_c=1,3$ for uncolored and colored
particles respectively. It is clear that the function $\,F(x)\,$
in (\ref{beta function}) equals one in the UV limit
$\,x\to\infty$, because this is required by the correspondence
between the mass-dependent renormalization and minimal subtraction
schemes at high energies. At the same time, in the opposite limit
$\,x\to 0$ this function has to vanish, because the non-decoupling
of the $\,\be_\La$-function would lead to the extremely fast
variation of the CC with the obvious untenable consequences in
cosmology. This means that in the power series expansion
$\,F(x)=F_0+xF_1+x^2F_2+...$, the coefficient $\,F_0\,$ is zero.
However, we do not have any reason to suppose that the other
coefficients of this expansion must be zero. It is easy to see
that assuming $\,F_1\neq 0\,$ leads to (\ref{m}). In fact, the
equality $\,F_1=0\,$ means that we impose one more constraint on
the CC, and this is not what we want, from the theoretical point
of view. So, let us suppose that $\,F_1\neq 0\,$ and hence that
the Eq. (\ref{m}) is true. Then the total $\,\beta$-function will
be the sum over all the fields and we reach the following
relation for the low-energy running of the cosmological constant:
\begin{eqnarray}\label{total}
\frac{d\La}{d\ln H}= \frac{F_1}{(4\pi)^2}
\sum_{i}\left(\frac{H^2}{m_i^2}\right) N_{J_i} m_i^4 \equiv\si
H^2 M^2\,,
\end{eqnarray}
where the parameter $\,\si\,M^2\,(\si=\pm 1)$ is defined by the
sum of all existing particles: light and heavy. Let us notice
that the leading contribution to (\ref{total}) are those of the
heaviest particles. Hence the details of the low-energy physics
has no impact on the possible infrared (IR) running of the CC.
This concerns, in particular, the non-perturbative effects of the
low-energy QCD and the higher loop contributions to the
$\,\be_\La$-function. By dimensional reasons and due to
covariance, these contributions will always fall into the same
form (\ref{total}). After all, this formula seems to have a
universal form and emerges from a vast class of QFT
models\footnote{It would be very interesting to look for the
analog of (\ref{total}) in string theory.}. Equation (\ref{total})
shows that the low-energy dynamics of the CC is completely defined
by the spectrum of the heaviest particles which become active only
at extremely high energies. We stress that, despite this looks
paradoxical, it is rather robust and the only one phenomenological
input which we used was the hypothesis $\,F_1\neq 0$. In fact, the
nontrivial role of the high energy spectrum for the low-energy
dynamics of the CC (\ref{total}) just means that we do not like to
introduce an extra unnecessary fine-tuning $\,F_1 = 0\,$ to the CC
problem.

\section{Impact for the early and late universes}

Equation (\ref{total}) includes a parameter $\,\si\,M^2\,$ which
we can not define from the present-day particle physics because it
may naturally involve Planckian-size masses of particles which are
inaccessible to all future accelerators. For the sake of
simplicity, we can just take $\,M^2=M_P^2$. Let us clarify that
this choice does not necessary mean that the relevant high energy
particles have the Planck mass. The mass of each particle may be
smaller than $M_P$, and the equality, or even the effective value
$M\gtrsim M_P$, can be achieved due to the multiplicities of these
particles. Finally, the sign $\,\si=\pm 1\,$ depends on whether
bosons or fermions dominate at the sub-Planck scale.

With these considerations in mind, our first observation is that
the natural value of the $\beta$-function (\ref{total}) at the
present time is
\begin{equation}
\Big|\beta_{\Lambda}\Big|
\,=\frac{c}{(4\pi)^2}\,M_P^2\,\cdot\,H_0^2\sim 10^{-47}\,GeV^4\,,
\label{beta Planck}
\end{equation}
where $c={\cal O}(1-10)$. Hence $\,\beta_{\Lambda}$ is very much
close to the experimental data on the CC\,\cite{SN}. This is
highly remarkable, because two vastly different and (in principle)
totally unrelated scales are involved to realize this
``coincidence'': $H_0$ and $M_P$, being these scales separated by
more than $60$ orders of magnitude! The resemblance between the
renormalization group eq. (\ref{total}) and the Friedmann equation
$H^2\sim \Lambda/M^2_P$ for the modern, CC-dominated, Universe
looks rather intriguing and is worth exploring. The RG equation
(\ref{total}) links the value of the CC with the one of the Hubble
parameter. The latter depends on the conformal factor of the
metric because of the dynamics of the matter-energy density. Let
us emphasize that the RG-based dependence between the CC and the
matter density means that we can essentially alleviate the
``new'' CC problem, because the coincidence is not directly
related to a particular cosmological epoch anymore! At the same
time, the dynamics of the CC may jeopardize the well-known
results in the Standard Cosmological Model, primarily for the
nucleosynthesis.

Let us consider the cosmological solution corresponding to
(\ref{total}). For the sake of simplicity we restrict our
consideration to the case of the conformally flat $\,k=0\,$ FLRW
metric. The more general formulas for an arbitrary $\,k\,$ can be
found in the parallel paper\,\cite{rgCC}. It proves useful to
solve for the CC and matter density in terms of the red-shift
variable $\,z\,$, defined as  $1+z=a_0/a$, where $a_0$ is the
present-day scale factor. Along with Eq. (\ref{total}) we shall
use the Friedmann equation
\begin{equation}
H^{2}\equiv \left( \frac{\dot{a}}{a}\right) ^{2}=\frac{8\pi G }{3}
\left( \rho +\Lambda\right)\,, \label{Friedmann}
\end{equation}
where $\rho$ is the matter-radiation density, and the energy
conservation law \beq
\frac{d\La}{dt}\,+\,\frac{d\rho}{dt}\,+\,3H\,(\rho+p)\,=\,0\,,
\label{conservation} \eeq where $p$ is the pressure. Since we need
to deal with both matter and radiation dominated regimes, it is
useful to solve the coupled system of differential equations
(\ref{total}), (\ref{Friedmann}) and (\ref{conservation}) using an
arbitrary equation of state $\,p=\al\rho$. The time derivative in
(\ref{conservation}) can be easily traded for a derivative in
$\,z\,$ via $d/dt=\,-\,H\,(1+z)\,d/dz$. The solution is completely
analytical and takes the form
\begin{equation}
\rho(z;\nu) \,=\,\rho_0\,(1+z)^{r} \qquad\qquad {\rm and}
\label{rho}
\end{equation}
\begin{equation}
\Lambda(z;\nu)\,=\,\Lambda_0
\,+\,\frac{\nu}{1-\nu}\,\left[\,\rho(z;\nu)-\rho_0\,\right]\,,
\label{Lambda}
\end{equation}
where $\rho_0, \Lambda_0$ are the present day values of the matter
density and CC, and we have introduced the following notations:
\begin{equation}
\nu\,=\,\frac{\sigma\,M^2}{12\pi M_P^2} \,,\,\,\,\,\,\,\,\,\,\,
r\,=\,3\,(1-\nu)\,(\al+1) \,. \label{notations 1}
\end{equation}
In order to avoid confusion, we note that the above solution for
$\,\La(z;\nu)\,$ has no singularity in the limit $\,\nu\rightarrow
1$. Also, as expected, for $\nu\rightarrow 0$ we recover the
standard result for $\,\rho(z)\,$ with constant CC.

Consider the nucleosynthesis epoch when the radiation dominates
over the matter, and derive the restriction on the parameter
$\,\nu$. In the radiation-dominated regime, the solution for the
density (\ref{rho}) can be rewritten in terms of the temperature
and the number of effectively massless or relativistic degrees of
freedom,
\begin{eqnarray}\label{rhozR2}
\rho_R(T)=\frac{\pi^2\,g_{\ast}}{30}
\,T^{4}\,\left(\frac{T_0}{T}\right)^{4\nu}
\end{eqnarray}
with  $\,T_{0}\simeq 2.75\,K=2.37\times 10^{-4}\,eV\,$ being the
present CMB temperature.

It is easy to see that the size of the parameter $\,\nu\,$ gets
restricted, because for $\,\nu \geq 1\,$ the density of radiation
(in the flat case) would be the same or even below the one at the
present universe. Hence, in order not to be ruled out by the
nucleosynthesis, our model has to satisfy
\begin{equation}
|\,{\Lambda_R}\,/\,{\rho_R}\,| \,\simeq \,
|\,{\nu}\,/\,{(1-\nu)}\,| \,\simeq \, |{\nu}|\ll 1\,.
\label{ratioCCrho}
\end{equation}

A nontrivial range could e.g. be $\,\,0<|{\nu}|\leq 0.1$. Both
signs of $\nu$ are in principle allowed provided the absolute
value satisfies the previous constraint. Let us notice that, in
view of the definition (\ref{notations 1}), the condition $\nu\ll
1$ also means that $M\lesssim M_P$. Hence, the nucleosynthesis
constraint coincides with our general will to remain in the
framework of the effective approach. It is remarkable that the two
constraints which come from very different considerations, lead to
the very same restriction on the unique free parameter of the
model. The canonical choice $\,M^2=M_P^2$, corresponds to
\begin{equation}\label{nurange}
|\nu|=\nu_0\equiv\frac{1}{12\,\pi}\simeq  2.6\times 10^{-2}\,.
\end{equation}

\section{The running and the ``old'' CC problem}

One can observe, at this point, some relation between our RG
approach and the ``old'' CC problem. There are two leading ideas
concerning the solution of this problem. The first one supposes
the existence of an unknown symmetry characterizing the
fundamental theory such as (super)string or M-theory. It is
supposed that this symmetry (an unbroken supersymmetry is the
simplest example) must preclude the contributions of the virtual
particles to the CC, and thus reduce the order of the fine-tuning
or even make it unnecessary (see, e.g. \cite{wittenDM}). An
obvious difficulty is that the corresponding symmetry must take
place at the very low energy scale, most naturally in the remote
future when the matter density will become zero. Also, this means
that the information about this symmetry was somehow encoded into
the early Universe, before all the supposed phase transitions took
place. We remark that our formula (\ref{total}) is the first
explicit example of the possible relation between the Planck scale
and the cosmic scale physics.

The second way of thinking about the CC problem is to assume that
its solution cannot be attained from first principles, and that
one must resort to some sort of anthropic
hypothesis\,\cite{weinRMP,weinDM}. Both points of view have some
peculiarities. The difficulty of the ``symmetry approach'' is that
one needs this symmetry not at high, but at very low energy.
Hence, many candidate symmetries are useless or at least looks to
be so. In particular, this concerns supersymmetry which (if exists
at all!) should be broken at low energies. In turn, the anthropic
hypothesis may be interpreted as an indication to the existence,
at some instant in the past, of many universes with some random
distribution of the values of the CC. One can, e.g., associate the
existence of the numerous choices of the universes with the
indefiniteness of vacuum in string theory which is indeed the main
candidate theory and is supposed to solve all problems of physics
including, of course, the CC one.

It is obvious that the first (symmetry-based) sort of solution for
the CC problem corresponds to the zero value of the CC in a remote
future, when the density of matter $\,\rho_M\,$ will become
negligible due to the further expansion of the Universe. At the
same time, the anthropic solution does not imply this requirement,
because the choice of the vacuum is performed in the chaotic way
and we are presumably living in just one of those universes which
permit the comfortable discussion of the CC problem. Actually, one
can get some hint about which way is correct by just applying the
solution (\ref{Lambda}) with the purpose to see whether the CC
can tend to zero in the remote future. It is easy to solve the
corresponding equation in the flat case: \beq
\La(z\to-1)\,=\,\La_0\,-\,\frac{\nu}{1-\nu}\,\rho_0\,=\,0\,.
\label{far} \eeq  We arrive at the suggestive value $\,\nu=
\,\Omega_{\Lambda}^0$, where the present day estimate is
$\Omega_{\Lambda}^0\simeq 0.7$\,\cite{SN}. This value of $\nu$ is
smaller than one, but it implies a fairly large correction to
some standard laws of conventional FLRW cosmology. Whether we can
accept it or not is not obvious at present. However, if accepted,
then it would hint at the ``symmetry'' approach to the old CC
problem, in the sense that string theory itself could perhaps
provide that value of $\nu$ as a built-in symmetry requirement.
In Ref. \cite{BigOne} we test explicitly the cosmological laws
using experimental data and simulations. We find that even
thinking of $\nu$ in Eqs.(\ref{rho}, \ref{Lambda}) as a mere
phenomenological parameter that gauges the departure of these
laws from the conventional FLRW solution, the tolerance in
$\nu\neq 0$ is still remarkably high.

\section{CC running and SNAP/HST testing}

We now ask whether even the most obviously permitted values of the
parameter $\nu \ll 1$ may lead to observable consequences. The
remarkable answer is: yes. In order to see this, we consider the
``recent'' Universe characterized by the redshift interval $\,0 <
z\lesssim 2$, and evaluate some cosmological parameters which can
be, in principle, improved by the future observations, say by the
SNAP project and beyond HST\,\cite{SNAP}. The first relevant
exponent is the relative deviation $\delta_{\Lambda}(z;\nu)\equiv
({\Lambda}(z;\nu)-{\Lambda}_0)/{{\Lambda}_0} \,$ of the CC from
the constant value $\La_0$. One has to remember that the existing
estimates for the CC from the supernovae data\,\cite{SN}
correspond to the supernovae measurements at some $z=z_0$. Then,
using our solution (\ref{Lambda}) we obtain, in first order of
$\nu$,
\begin{eqnarray}
\delta_{\Lambda}(z;\nu)=\,\frac{\nu
\,\Omega_M^0}{\Omega_{\Lambda}^0} \left[(1+z)^3-(1+z_0)^3\right].
\label{deviation Lambda}
\end{eqnarray}
Taking $z_0\simeq 0.5$  (the approximate central value of the
sample of high redshift supernovae from\,\cite{SN}), with
$\,\Omega_M^0=0.3\,$ and $\Omega_{\Lambda}^0=0.7$, and
$\nu=\nu_0$, we find e.g. $\delta_{\Lambda}(z=1.5;\nu_0)=14\%$. In
general, the strong cubic $z$-dependence in
$\,\delta_{\Lambda}(z;\nu)\,$ should manifest itself in the future
CC observational experiments where the range $\,z\gtrsim 1$ will
be tested. It is important to emphasize that $\,\nu\,$ is the
unique arbitrary parameter of this model for a variable CC.
Therefore, the experimental verification of the above formula must
consist in: {\it i)} pinning down the sign and value of the
parameter $\,\nu$; and {\it ii)} fitting that formula to the
experimental data\,\footnote{For alternative RG scenarios, in the
context of quantum gravity, see \cite{ReuterBonanno} and
references therein.}.

Next we present the relative deviation of the square of the Hubble
parameter $\,H^2(z,\nu)\,$ with respect to the conventional one
$\,H^2(z,\nu=0)$. Again we just quote the flat case. The resulting
deviation $\delta H^2(z;\nu)\equiv
(H^2(z;\nu)-H^2(z;0))/{H^2(z;0)}$ is
\begin{eqnarray}\label{deltaH}
\delta H^2=-\nu\Omega_M^0\frac{1+(1+z)^3
\left[3\ln(1+z)-1\right]}{1+\Omega_M^0\left[(1+z)^3-1\right]}.
\end{eqnarray}
Equation (\ref{deltaH}) gives the leading quantum correction to
the Hubble parameter (\ref{Friedmann}) when the renormalization
effects in (\ref{rho}) and (\ref{Lambda}) are taken into account.
Notice that $\delta H^2(0;\nu)=0$, because for all $\nu$ we have
the same initial conditions. Then for $z\neq 0$ we have e.g.\,
$\delta H^2(1.5;\nu_0)\simeq\,-4.2\%$ and $\delta
H^2(2;\nu_0)\,\simeq\,-5.7\%$. For larger $\nu$, we get quite
sizeable effects like $\delta H^2(z;0.1)\simeq\,-16\%$ and
$-21\%\,$ for $z=1.5$ and $z=2$ respectively.

The last exponent of interest that we wish to remark here makes
use of our previous results for  $\Lambda(z;\nu)$ and $H(z;\nu)$.
Then we can compute the relative deviation of the renormalized
cosmological constant parameter
$\Omega_{\Lambda}(z;\nu)=8\pi\,G\Lambda(z;\nu)/3H^2(z;\nu)$ at
redshift $\,z$ with respect to the standard one,
$\Omega_{\Lambda}(z;0)$. At leading order,
$$
\delta {\Omega_{\Lambda}}(z;\nu)= \frac{\Omega_{\Lambda}(z;\nu)
-\Omega_{\Lambda}(z;0)}{\Omega_{\Lambda}(z;0)}\,=
$$$$
=\,\nu\left[\frac{{\Omega}_{M}^0(1+z)^3-1}{\Omega_{\Lambda}^0}\,+
\right.
$$
\begin{equation}
\label{deviOmega1}
\left.
 +\,\frac{1+3{\Omega}_{M}^0(1+z)^3\,\ln(1+z)}
{\Omega_{\Lambda}^0+{\Omega}_{M}^0(1+z)^3}\right].
\end{equation}
Again $\,\,\delta\Omega_{\Lambda}(0;\nu)=0$, as it should.
Moreover, the deviation has the two expected limits for the
infinite past and future, viz.
$\delta\Omega_{\Lambda}(\infty;\nu)=\infty$ and
$\delta\Omega_{\Lambda}(-1;\nu)=0$.  Notice that even for $\nu$ as
small as $\nu_0$, (\ref{nurange}), there is a sizeable $20\%$
increase of $\Omega_{\Lambda}$  at redshift  $z=1.5$ -- reachable
by SNAP. For $\nu=2\,\nu_0$ the increase at $z=1.5$ is huge,
$40\%$. For $\nu<0$ the effects go in the opposite direction. If
some future experiments can reach the far $z=2$ region  with
enough statistics, the effects on $\Omega_{\Lambda}(z)$ are even
more dramatic. At present $\Omega_{\Lambda}^0$ has been determined
at roughly $10\%$ from both supernovae and CMB measurements, and
in the future SNAP will pin $\Omega_{\Lambda}$ down to within $\pm
0.05$\,\cite{SNAP}. The previous numbers show that for $z\gtrsim
1$, the cosmological quantum corrections can be measured already
for a modest $\nu\gtrsim 10^{-2}$.  A complete numerical analysis
of this kind of FLRW models, including both the flat and curved
space cases, together with a detailed comparison with the present
and future Type Ia supernovae data, will be presented
elsewhere\,\cite{BigOne}. \vskip 6mm

\section{Conclusions}

We have exemplified the possible running of the CC at the present
cosmic scale due to the renormalization group and the smooth
decoupling of the massive fields at low energies, assuming the
Hubble parameter $H$ as the RG scale. A time dependence of the CC
may therefore be achieved without resorting to scalar fields
mimicking the cosmological term. It turns out that the $\be_\La$
function has just one arbitrary parameter $\,\nu<1$. For
$\,\nu\ll 1$, we insure the absence of the transplanckian
energies and also consistency of the CC with the nucleosinthesis
calculations. However, larger values of $\,\nu$ cannot be
completely excluded at present\,\cite{BigOne}. For example, if
$\,\nu=\Omega_{\Lambda}^0$, a flat universe would have exactly
zero CC in the infinite future. This would open the possibility
that a symmetry requirement, e.g. within M-theory, could avoid
the embarrassing event horizon problem in this string framework
where an asymptotic positive CC is not welcome. In fact, our
model with running CC could represent the effective behavior of
many high energy theories. Last, but not least, there are
excellent prospects for testing this RG cosmological model in the
future by SNAP and the upgraded HST experiments.

\textit{Acknowledgments}: The authors are thankful to E.V. Gorbar,
C. Espa\~{n}a-Bonet and P. Ruiz-Lapuente for fruitful discussions.
I.Sh. has been partially supported by FAPEMIG and CNPq. J.S. has
been supported in part by MECYT and FEDER  and also by the Dep.
de Recerca de la Generalitat de Catalunya under the BE 2002
program. J.S. thanks the warm hospitality at the Dep. de Fisica
UFJF.

\providecommand{\href}[2]{#2}

\end{document}